\documentclass[preprint,aps]{revtex4}
\usepackage{mathrsfs}
\usepackage{graphicx}
\usepackage{multirow}
\usepackage{makecell}
\usepackage{booktabs}
\usepackage{bm}

\begin{document}

\title{Long-Time Magnetic Relaxation in Antiferromagnetic Topological Material EuCd$_2$As$_2$}

\author{Yang Wang$^{1,2\sharp}$, Cong Li$^{1,2\sharp}$, Yong Li$^{1,2\sharp}$, Xuebo Zhou$^{1,2}$, Wei Wu$^{1,2}$, Runze Yu$^{1,2}$, Jianfa Zhao$^{1,2}$, Chaohui Yin$^{1,2}$, Youguo Shi$^{1,2,3}$, Changqing Jin$^{1,2,3}$, Jianlin Luo$^{1,2,3}$, Lin Zhao$^{1,2,3}$, Tao Xiang$^{1,2,4}$, Guodong Liu$^{1,2,3*}$ and  X. J. Zhou$^{1,2,3,4*}$
}

\affiliation{
\\$^{1}$Beijing National Laboratory for Condensed Matter Physics, Institute of Physics, Chinese Academy of Sciences, Beijing 100190, China
\\$^{2}$University of Chinese Academy of Sciences, Beijing 100049, China
\\$^{3}$Songshan Lake Materials Laboratory, Dongguan 523808, China
\\$^{4}$Beijing Academy of Quantum Information Sciences, Beijing 100193, China
\\$^{\sharp}$These people contributed equally to the present work.
\\$^{*}$Corresponding authors: gdliu$\_$ARPES@iphy.ac.cn and XJZhou@iphy.ac.cn
}

\date{May 21, 2021}


\begin{abstract} 
Magnetic topological materials have attracted much attention due to the correlation between topology and magnetism. Recent studies suggest that EuCd$_2$As$_2$ is an antiferromagnetic topological material. Here by carrying out thorough magnetic, electrical and thermodynamic property measurements, we discover a long time relaxation of the magnetic susceptibility in EuCd$_2$As$_2$. The (001) in-plane magnetic susceptibility at 5 K is found to continuously increase up to $\sim$10\% over the time of $\sim$14 hours. The magnetic relaxation is anisotropic and strongly depends on the temperature and the applied magnetic field. These results will stimulate further theoretical and experimental studies to understand the origin of the relaxation process and its effect on the electronic structure and physical properties of the magnetic topological materials.
\end{abstract}

\maketitle




Searching for new topological states with novel exotic properties has become an important subject in condensed-matter physics\cite{MZHasan_RMP2010,XLQi_RMP2011,HMWeng_MRSBull2014,NPArmitage_RMP2018,HWatanabe_SciAdv2018,YTokura_NatRevPhys2019,TZhang_Nature2019,MGVergniory_Nature2019,MMOtrokov_Nature2019,FTang_Nature2019,YFXu_Nature2020}. Magnetic topological materials, as a new category of topological materials, have attracted much recent attention\cite{RYu_Sci2010,RSKMong_PRB2010,YLChen_Sci2010,XGWan_PRB2011,AABurkov_PRL2011,GXu_PRL2011,CZChang_Science2013,PZTang_NP2016,MHirschberger_NatMater2016,EEmmanouilidou_PRB2017,MMOtrokov_JETP2017,JWang_arxiv2017,HYang_NJP2017,KKuroda_NatMater2017,SNie_PNAS2017,KKim_NatMater2018,EKLiu_NP2018,QWang_NC2018,DFLiu_Science2019,MZHasan_Science2019,SBorisenko_NC2019,YFXu_PRL2019,IBelopolski_Science2019,JZhou_NPJCM2019,MMOtrokov_Nature2019,YGong_CPL2019,DZhang_PRL2019,HLi_PRX2019,YDeng_Science2020,JGe_NationSciRev2020,YFXu_Nature2020,XLQi_PRB2008,FWilczek_PRL1987,AMEssin_PRL2009,KNomura_PRL2011,TMorimoto_PRB2015,JWang_PRB2015,MMogi_NatMater2017,MMogi_SciAdv2017,DXiao_PRL2018,CLiu_NatMat2020,WABenalcazar_Science2017,FSchindler_SciAdv2018,LSmejkal_NP2018}. Compared with non-magnetic topological materials, the introduction of magnetism can break the time-reversal symmetry and produce new quantum topological phases such as the quantum anomalous Hall state\cite{RYu_Sci2010,CZChang_Science2013}, the axion state\cite{XLQi_PRB2008,FWilczek_PRL1987,AMEssin_PRL2009,KNomura_PRL2011,TMorimoto_PRB2015,JWang_PRB2015,MMogi_NatMater2017,MMogi_SciAdv2017,DXiao_PRL2018,YFXu_PRL2019,CLiu_NatMat2020} and even high-order topological phase\cite{MSitte_PRL2012,FZhang_PRL2013,WABenalcazar_Science2017,FSchindler_SciAdv2018}. These new physical states greatly broaden the field of the topology research as well as have a great application potential in many fields such as spintronics and topological quantum computing\cite{LSmejkal_NP2018,YTokura_NatRevPhys2019}. Much efforts have been devoted to looking for intrinsic magnetic topological materials because they can reduce disorder effect and achieve quantized states at higher temperature\cite{YTokura_NatRevPhys2019,YDeng_Science2020,JGe_NationSciRev2020}. Although some intrinsic magnetic topological materials have been predicted, very few have been experimentally verified\cite{MHirschberger_NatMater2016,EEmmanouilidou_PRB2017,HYang_NJP2017,KKuroda_NatMater2017,SNie_PNAS2017,GYHua_PRB2018,KKim_NatMater2018,EKLiu_NP2018,QWang_NC2018,IBelopolski_Science2019,DFLiu_Science2019,MZHasan_Science2019,SBorisenko_NC2019,JZhou_NPJCM2019,MMOtrokov_Nature2019,YGong_CPL2019,DZhang_PRL2019,HLi_PRX2019,YFXu_Nature2020}.

Recent work has shown that EuCd$_2$As$_2$ is an interesting magnetic topological material where multiple topological states can be realized by considering the magnetic structure and crystal symmetries\cite{GYHua_PRB2018,LLWang_PRB,CWNiu_PRB,JZMa_AM,JZMa_SA}. Various magnetic structures in EuCd$_2$As$_2$ are proposed\cite{GYHua_PRB2018,Krishna_PRB}, experimentally detected\cite{Artmann_Chem,Schellenberg_Chem,HPWang_PRB,MCRahn_PRB,JZMa_SA,JZMa_AM,Soh_PRB2020} and manipulated\cite{Soh_PRB2019,Jo_PRB,Sanjeewa_PRB,YXu_PRB}. When EuCd$_2$As$_2$ is in the inter-layer antiferromagnetic state, it hosts only one pair of Dirac points at the Fermi level\cite{GYHua_PRB2018}. Inducing ferromagnetism along the c axis can generate a single pair of Weyl points that provides an ideal case to study the unique properties of Weyl semimetals\cite{LLWang_PRB,JZMa_SA,Soh_PRB2019}. Different magnetic structures combined with rotational or inversion symmetry breaking in EuCd$_2$As$_2$, can result in a plethora of other topological states including antiferromagnetic topological insulators\cite{JZMa_AM}, a triple-point magnetic topological semimetal\cite{GYHua_PRB2018} and a quantum anomalous Hall insulator\cite{CWNiu_PRB}. EuCd$_2$As$_2$ has become a fertile playground for studying the interplay between magnetism and topology and the magnetic structure plays a key role in dictating its topological properties.

In this paper, we carried out thorough magnetic, electrical and thermodynamic property measurements on EuCd$_2$As$_2$. We discover a long time relaxation of the (001) in-plane magnetic susceptibility in EuCd$_2$As$_2$. The in-plane magnetic susceptibility at 5 K is found to continuously increase up to $\sim$10\% over the time of $\sim$14 hours. We also find that this slow magnetic relaxation does not have an obvious effect on the bulk properties including the crystal structure, the resistivity and the specific heat. Our unexpected finding of the slow magnetic relaxation process in EuCd$_2$As$_2$ will simulate further studies to understand its microscopic origin and its manifestation in other physical properties.

Figure 1 shows the crystal structure and the magnetic structure of EuCd$_2$As$_2$. EuCd$_2$As$_2$ crystallizes in a trigonal structure with a P$\bar3$m1 space group (no.164)\cite{Artmann_Chem}. The double-corrugated Cd-As layers are sandwiched between the hexagonal Eu layers, as shown in Fig. 1a. It has been shown that EuCd$_2$As$_2$ undergoes a magnetic transition at 9.5 K from a paramagnetic state to an antiferromagnetic (AFM) state\cite{Artmann_Chem}. The magnetism of EuCd$_2$As$_2$ originates from the localized 4f electrons of the Eu atoms in the Eu layers. According to the theoretical calculations\cite{Krishna_PRB,GYHua_PRB2018}, EuCd$_2$As$_2$ may have A-type antiferromagnetic structure with in-plane or out-of-plane magnetic moments at low temperatures, as shown in Fig. 1b with three possible A-type AFM  structures on Eu sites with magnetic moments along $\bf{b}$, $\bf{c}$ and $\bf{x}$ directions. While some measurements suggest that the magnetic moment is along the c direction\cite{HPWang_PRB}, most experiments support that it is located within the a-b plane\cite{MCRahn_PRB,JZMa_SA,JZMa_AM,Jo_PRB,Sanjeewa_PRB}. 

High quality single crystals of $\mathrm{EuCd_2As_2}$ were grown by Sn flux method\cite{HPWang_PRB}. The single crystals were characterized by Laue diffraction and X-ray diffraction (XRD). Fig. 1c shows the Laue pattern of EuCd$_2$As$_2$ crystal for the (001) plane. It shows sharp diffraction spots with a three-fold symmetry. The X-ray diffraction was measured by using a rotating anode X-ray diffractometer with Cu $K_\alpha$ radiation ($\lambda$ = 1.5418 $\AA$). All the observed peaks can be attributed to the Miller indices (00$l$) and the extracted lattice constant c=7.322 $\AA$.

We further characterized our samples of EuCd$_2$As$_2$ single crystal by powder X-ray diffraction measured at different temperatures. To this end, the powder sample was obtained by grinding the  EuCd$_2$As$_2$ single crystals. Fig. 2a shows the XRD patterns of the EuCd$_2$As$_2$ powder at different temperatures. All the observed peaks can be attributed to EuCd$_2$As$_2$ with a trigonal structure in P$\bar3$m1 space group, as marked on the 300 K data except for the two peaks (marked by asterisks in Fig. 2a) that are from the copper sample holder. Over the temperature range between 12\,K and 300\,K, the XRD patterns exhibit no obvious change in terms of the number of the observed peaks, consistent with the absence of structure transition in this temperature range. The peak position shows a clear variation with temperature, as exemplified by the (110) peak shown in Fig. 2b, which indicates that the lattice constants change with temperature. The structural refinement of the observed XRD patterns in Fig. 2a gives the lattice constants a(b) and c as different temperatures, as shown in Fig. 2c and 2d, respectively. From 300\,K to 12\,K, the lattice constants a and c decrease by 0.22\% and 0.34\%, respectively.

Figure 3 shows magnetic susceptibility measurements of EuCd$_2$As$_2$. The magnetic measurement was carried out by using magnetic property measurement system-3 (MPMS-3). The temperature-dependent magnetic susceptibility is measured in two modes (zero-field-cooled, ZFC, and field-cooled, FC) under different applied magnetic fields parallel (Fig. 2a) and perpendicular (Fig. 2b) to the (001) plane. The residual magnetic field at the sample position is obtained by measuring and renormalizing the magnetic susceptibility at a high temperature in the paramagnetic state under different magnetic fields. The magnetic field shown in this work is corrected after considering the residual field. First, the magnetic susceptibility displays an obvious antiferromagnetic transition at T$_N\sim$9.3\,K in all the measurements (marked by red arrows in the upper-left insets of Fig. 3a and 3b). Second, the magnetic susceptibility measured under different magnetic fields bifurcates at a temperature (T$_c$) higher than T$_N$; it is $\sim$16\,K for the in-plane measurement (marked by black arrow in the upper-left inset of Fig. 3a) and $\sim$14\,K for the out-of-plane measurement (marked by black arrow in the upper-left inset of Fig. 3b). This is consistent with the previous measurements and such behaviors are attributed to a ferromagnetic transition\cite{Artmann_Chem}. Third, in the paramagnetic state at high temperature, the magnetic susceptibility is the same when measured under different magnetic fields. It also satisfies the Curie-Weiss law $\chi=C/(T-\theta)$, as shown in the upper-right insets of Fig. 3a and 3b. Fourth, in the ferromagnetic or antiferromagnetic state at low temperature, the magnetic susceptibility differs when measured in different modes between ZFC and FC measurements. The difference is significant when the applied magnetic field is small and gets suppressed when the applied field is large. Fifth, the magnetic susceptibility exhibits a strong anisotropy; the (001) in-plane susceptibility $\chi_{ab}$ at the N\'eel temperature T$_N$ in Fig. 3a is nearly 4 times the out-of-plane susceptibility $\chi_{c}$ in Fig. 3b. It indicates that the magnetic moments lie mainly in the (001) plane. Our magnetic measurements are consistent with the previous results\cite{Artmann_Chem,Schellenberg_Chem,MCRahn_PRB}. The in-plane A type AFM structure was directly determined by resonant elastic x-ray scattering in Ref. \cite{MCRahn_PRB}. Therefore, our EuCd$_2$As$_2$ samples should have the magnetic ground state because the samples we used are the same as those used in Ref. \cite{MCRahn_PRB} and they come from the same sample provider. Our measured temperature- and magnetic-field- dependent magnetization curves are consistent with those in Ref. \cite{MCRahn_PRB}.

The electrical resistivity measurement of EuCd$_2$As$_2$ was carried out by the standard four-probe method (lower-right inset in Fig. 4a) using a physical property measurement system-14 (PPMS-14). Fig. 4a shows the temperature dependence of the (001) in-plane resistivity measured under different applied magnetic fields normal to the (001) plane. Without the applied magnetic field, the resistivity exhibits a metallic-like behavior between 50\,K and 300\,K. Below 50\,K, the resistivity shoots up with decreasing temperature, reaches a maximum at $\sim$9.3\,K and drops again with further decrease of the temperature. The occurrence of the resistivity peak is closely related to the antiferromagnetic transition at T$_N\sim$9.3\,K. Upon applying the magnetic field, the resistivity peak drops in its magnitude, broadens in its peak width, and shifts to higher temperature in its peak position. These results are consistent with the previous measurements\cite{MCRahn_PRB}. We further find that the resistivity peak temperature increases linearly with the applied magnetic field, as shown in the upper inset in Fig. 4a. It is clear in Fig. 4a that, at low temperature, the resistivity of EuCd$_2$As$_2$ exhibits a strong dependence on the applied magnetic field. To investigate the magnetoresistance effect quantitatively, we measured the magnetic field dependent resistivity at different temperatures with the magnetic field perpendicular to the (001) plane, as shown in Fig. 4b. At 5\,K in the antiferromagnetic state, the resistivity initially rises rapidly with the applied magnetic field, reaches a maximum at $\sim$0.1\,T, then drops precipitously till about 0.5\,T before it levels off at higher magnetic field. The dominant resistivity change is confined within a narrow range of the applied magnetic field. Right at the antiferromagnetic transition temperature of 9.3\,K, the resistivity starts to drop right after the magnetic field is applied and then the resistivity change spreads to a wider range of the magnetic field. At 15\,K that is above the antiferromagnetic transition temperature, the resistivity increases with the applied magnetic field and reaches a maximum at $\sim$0.65\,T, then drops gradually with the magnetic field. The magnitude of the resistivity change at 15\,K is obviously smaller than those at 5\,K and 9.3\,K, and the change spreads over a much wider range of the applied magnetic field. The strongest negative magnetoresistance signal appears at 9.3\,K which indicates that the magnetoresistance behaviors of EuCd$_2$As$_2$ are intimately related with its magnetic structure.

We also carried out measurements of the specific heat, the Seebeck coefficient and the thermal conductivity on EuCd$_2$As$_2$. These measurements were performed in the physical property measurement system-9 (PPMS-9). Fig. 5a shows the measured specific heat as a function of temperature. The specific heat drops from 300\,K to 13\,K, reaches a minimum at around 13\,K and exhibits a peak at 9.3\,K that is apparently related with the antiferromagnetic transition. Fig. 5b shows the Seebeck coefficient of EuCd$_2$As$_2$ as a function of temperature measured under two different magnetic fields. The Seebeck coefficient is positive over the entire temperature range, indicating that the dominant charge carriers of EuCd$_2$As$_2$ is hole-like. It decreases with decreasing temperature between 300\,K and $\sim$40\,K and exhibits a sharp peak at 9.3\,K (black curve in Fig. 5b) which is suppressed when a high magnetic field is applied (red curve in Fig. 5b). Fig. 5c shows the temperature-dependent thermal conductivity of EuCd$_2$As$_2$ measured under two different magnetic fields. It increases with the decrease of the temperature and exhibits a sharp peak at 9.3\,K (black curve in Fig. 5c). This peak is slightly suppressed when a magnetic field of 8\,T is applied (red curve in Fig. 5c). Our results indicate that the transport and thermodynamic properties of EuCd$_2$As$_2$ are intimately related to its magnetic structure at low temperature.

Unexpectedly, we find that EuCd$_2$As$_2$ exhibits a long-time relaxation in its magnetic susceptibility. Fig. 6 shows the variation of the magnetic susceptibility with time after the sample is quickly cooled down from the room temperature to 5\,K and the magnetic susceptibility is immediately measured under a magnetic field of 3.7\,Oe. When the magnetic field is parallel to the (001) plane (Fig. 6a), the in-plane magnetic susceptibility $\chi_{ab}$ shows a quick increase initially, followed by a gradual increase later on. Over a period of $\sim$14 hours, it increases by about $\sim$10\% (black curve in Fig. 6c). The out-of-plane magnetic susceptibility $\chi_c$ (Fig. 6b) also exhibits an obvious increase with the relaxation time. The relative change over the same period time of $\sim$14 hours is about $\sim$3\% (red curve in Fig. 6c) which is much smaller than that of the in-plane magnetic susceptibility. We note that, even though the magnitude of the relative change is different, the curve shape of the in-plane and out-of-plane magnetic susceptibilities is identical when they are scaled at the data points of long relaxation time. When trying to fit the data with an usual formula involving one relaxation time $\tau$: $a+b*exp(-t/\tau)$ (blue curve in Fig. 6c), we find that it is impossible to match the data. When using a formula with two relaxation time $\tau_1$ and $\tau_2$: $a+b*exp(-t/\tau_1)+c*exp(-t/\tau_2)$ (red and green curves in Fig. 6c), we find that the fitted data can capture the main characteristic of the measured data although there are still some deviations. These results indicate that the slow magnetic relaxation process in EuCd$_2$As$_2$ is unusual and may involve multiple time scales.

In order to check how the magnetic relaxation process in EuCd$_2$As$_2$ depends on the temperature, we carried out the magnetic relaxation experiments at three typical temperatures, 5\,K below the N\'eel temperature, 9.3\,K at the N\'eel temperature and 15\,K above the N\'eel temperature. Fig. 7(a-c) shows the relaxation results at these three temperatures under an applied magnetic field of 3.7\,Oe. The relative change of the magnetic susceptibility at the three temperatures is summarized in Fig. 7d. It is found that the magnetic relaxation depends strongly on the sample temperature. While the magnetic relaxation is significant at low temperature (5\,K), it gets weaker with increasing temperature and becomes nearly negligible at 15\,K. Fig. 7e shows the comparison of the magnetic susceptibility before and after the relaxation at different temperatures measured under a magnetic field of 3.7\,Oe. The magnetic property of EuCd$_2$As$_2$ has been changed after the relaxation process and the effect is most prominent at low temperature below the antiferromagnetic transition at 9.3\,K (Fig. 7e). Such an effect is observable when the magnetic field used to determine the magnetic susceptibility is small, such as 3.7\,Oe used in Fig. 7e. The effect becomes invisible when the magnetic field is large, as shown in Fig. 7f. When the magnetic field is 3.7 Oe (Fig. 7e), the magnetic susceptibility shows a clear dependence on the relaxation temperature in the ZFC measurement mode but it exhibits little difference when measured in the FC mode. When the magnetic field is 500 Oe (Fig. 7f), the magnetic susceptibility shows little dependence on the relaxation temperature in both the FC and ZFC modes. These results indicate that the magnetic relaxation occurs in the magnetic state; the lower is the sample temperature, the stronger is the magnetic relaxation.

We also checked the effect of the applied magnetic field on the magnetic relaxation process in EuCd$_2$As$_2$. Fig. 8(a-d) shows the relaxation results under different magnetic fields of 3.7\,Oe (a), 20\,Oe (b), 100\,Oe (c) and 500\,Oe (d) measured at 5\,K. The relative change of the magnetic susceptibility under these four magnetic fields is summarized in Fig. 8e. It is clear that the magnetic relaxation also depends sensitively on the applied magnetic field. When the magnetic field is small like 3.7\,Oe, the relaxation process is strong with a relative magnetic susceptibility change up to 10\% over a period of time 14 hours (red curve in Fig. 8e). When the magnetic field gets larger, the relaxation process becomes weaker (blue and green curves in Fig. 8e). The relaxation process is strongly suppressed when the applied magnetic field is large like 500\,Oe (yellow curve in Fig. 8e). The magnetic property of EuCd$_2$As$_2$ relaxed at different magnetic field is changed and the change is obvious at low temperature in the magnetic state (Fig. 8f). Again, this change is observable when the magnetic field used to determine the magnetic susceptibility is small, such as 3.7\,Oe in Fig. 8f, but becomes invisible when the magnetic field is large, as shown in Fig. 8g. These results indicate that for the magnetic relaxation in EuCd$_2$As$_2$ to occur in the magnetic state, the smaller is the applied magnetic field, the stronger is the relaxation effect.

The magnetic relaxation process we have observed indicates that the magnetic structure changes with time in the bulk EuCd$_2$As$_2$. It is natural to ask whether the bulk crystal structure and physical properties also vary with time in EuCd$_2$As$_2$. To check on the resistivity change, we cooled down quickly the sample from the room temperature to 5\,K and measured the in-plane resistivity at 5\,K after different time (Fig. 9a). Little change ($\textless$0.5\%) is observed in the resistivity over a period of 78 hours (Fig. 9a). In order to check the change of the specific heat, we cooled down the sample quickly from the room temperature to 9.3\,K and measured the specific heat at different time. We also did the same measurement by cooling down the sample to 5\,K. As shown in Fig. 9b, at both temperatures, the specific heat keeps nearly constant over a period of $\sim$20 hours. To check the effect of the magnetic relaxation on the crystal structure, we carried out powder XRD measurement at different time after the sample is quickly cooled down from the room temperature to 12\,K, as shown in Fig. 9c. Over a period of 18 hours, the measured XRD patterns are identical and no obvious change is found with time (Fig. 9c).

It is the first time to observe the slow magnetic relaxation process in EuCd$_2$As$_2$. The relaxation is unusual because the process is slow that can last for hours and the effect is significant that the change of the magnetic susceptibility with time can reach $\sim$10\%. The sluggish magnetic relaxation in EuCd$_2$As$_2$ is rather similar to that recently discovered in (MnBi$_2$Te$_4$)(Bi$_2$Te$_3$)$_n$\cite{JZWu_AdvMater2020}, in terms of the bifurcation between the ZFC and FC measurements (Fig. 3), the relaxation curve shape that can not be described by a simple Arrhenius law (Fig. 6), and the dependence of the relaxation on the applied magnetic field (Fig. 8). The slow magnetic relaxation observed in (MnBi$_2$Te$_4$)(Bi$_2$Te$_3$)$_n$ with n$\ge$2 is considered as a signature of two-dimensional magnet where the inter-layer magnetic coupling is approaching zero\cite{JZWu_AdvMater2020}. In EuCd$_2$As$_2$, the inter-layer magnetic coupling is strong that gives rise to the antiferromagnetic transition below 9.3\,K. Therefore, our finding of the similar relaxation behaviors in EuCd$_2$As$_2$ strongly indicates that two dimensionality is not a necessary factor to induce the slow magnetic relaxation process. As in (MnBi$_2$Te$_4$)(Bi$_2$Te$_3$)$_n$, the microscopic origin of the slow magnetic relaxation remains to be investigated as it involves complex interactions of many factors such as the inter-layer coherent spin rotation and intra-layer domain wall movement\cite{JZWu_AdvMater2020}. Our new finding of the magnetic relaxation in EuCd$_2$As$_2$ can provide new insight in uncovering the relaxation mechanism in these magnetic topological materials.

In summary, by carrying out magnetic, electrical transport and thermodynamic measurements on EuCd$_2$As$_2$, a long-time relaxation of the magnetic susceptibility in EuCd$_2$As$_2$ is observed for the first time. It is anisotropic that the (001) in-plane relaxation is stronger than the out-of-plane case. The effect gets stronger with decreasing temperature in the magnetic state. It is prominent under low magnetic field and gets suppressed when the applied magnetic field is large. These observations will stimulate further theoretical and experimental studies to understand the origin of the relaxation process and its effect on the electronic structure and physical properties of the magnetic topological materials.\\


\vspace{3mm}

\noindent {\bf Acknowledgement}\\
We thank Changjiang Yi for the EuCd$_2$As$_2$ crystal growth, single XRD measurement and the discussion in data process. This work is supported by the National Key Research and Development Program of China (Nos. 2016YFA0300600 and 2018YFA0305600), the National Natural Science Foundation of China (No. 11974404), the Strategic Priority Research Program (B) of the Chinese Academy of Sciences (No. XDB33000000), and the Youth Innovation Promotion Association of CAS (No. 2017013).

\vspace{3mm}


\newpage

\begin{figure*}[tbp]
\begin{center}
\includegraphics[width=1\columnwidth,angle=0]{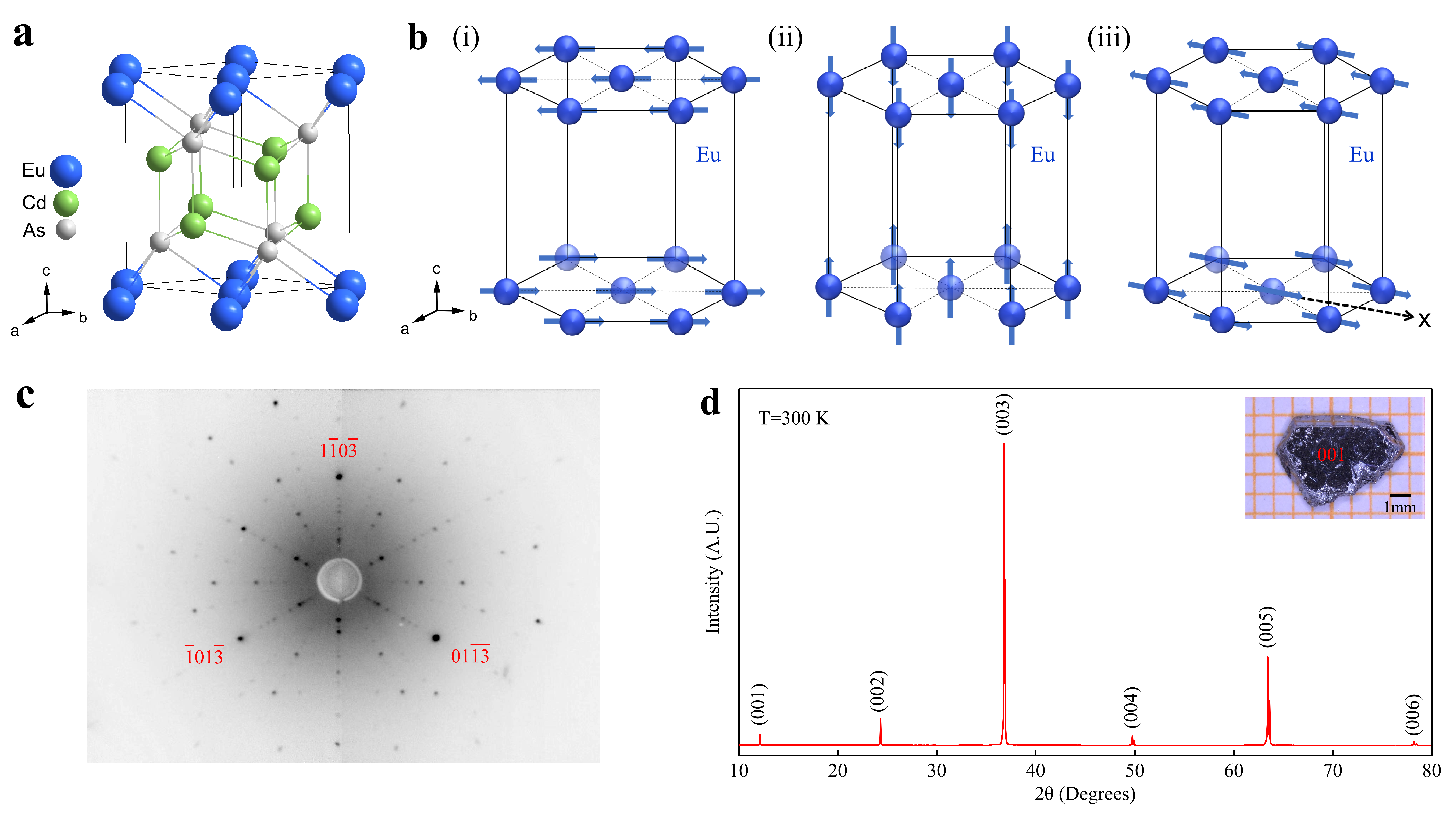}
\end{center}
\caption{\textbf{Crystal structure, magnetic structure and crystal characterization of EuCd$_2$As$_2$.} (a) Crystal structure of EuCd$_2$As$_2$. (b) Proposed possible magnetic structures in EuCd$_2$As$_2$. They represent A-type antiferromagnetic structure on Eu sites with the magnetic moment along $\bf{b}$ (i), $\bf{c}$ (ii) or $\bf{x}$ (iii) directions\cite{Krishna_PRB}. (c) Laue diffraction pattern of EuCd$_2$As$_2$ single crystal from (001) cleavage plane. (d) Single crystal XRD pattern of EuCd$_2$As$_2$ measured at 300K. The photo of EuCd$_2$As$_2$ crystal with (001) plane is shown in the upper-right inset.
}
\end{figure*}

\begin{figure*}[tbp]
\begin{center}
\includegraphics[width=1\columnwidth,angle=0]{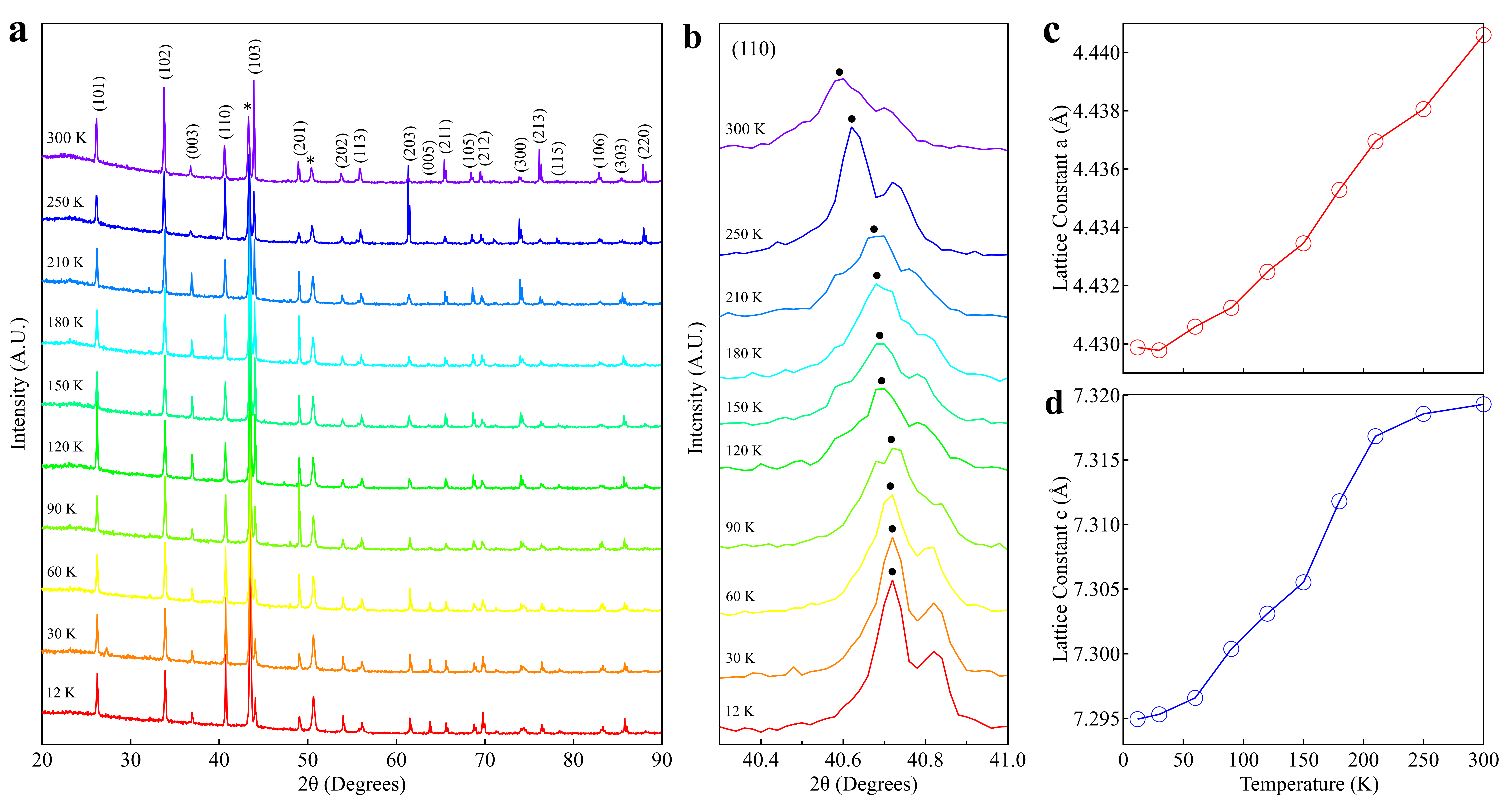}
\end{center}
\caption{\textbf{Temperature-dependent powder XRD measurements of EuCd$_2$As$_2$}. (a) Powder XRD patterns of EuCd$_2$As$_2$ measured at different temperatures. All the observed peaks can be indexed as shown on the 300 K data except for the two peaks marked by asterisks which can be attributed to copper from the sample holder. (b) Zoom-in (110) peaks at different temperatures. (c-d) Temperature dependence of the lattice constants a (c) and c (d) extracted from (a).
}
\end{figure*}

\begin{figure*}[tbp]
\begin{center}
\includegraphics[width=1\columnwidth,angle=0]{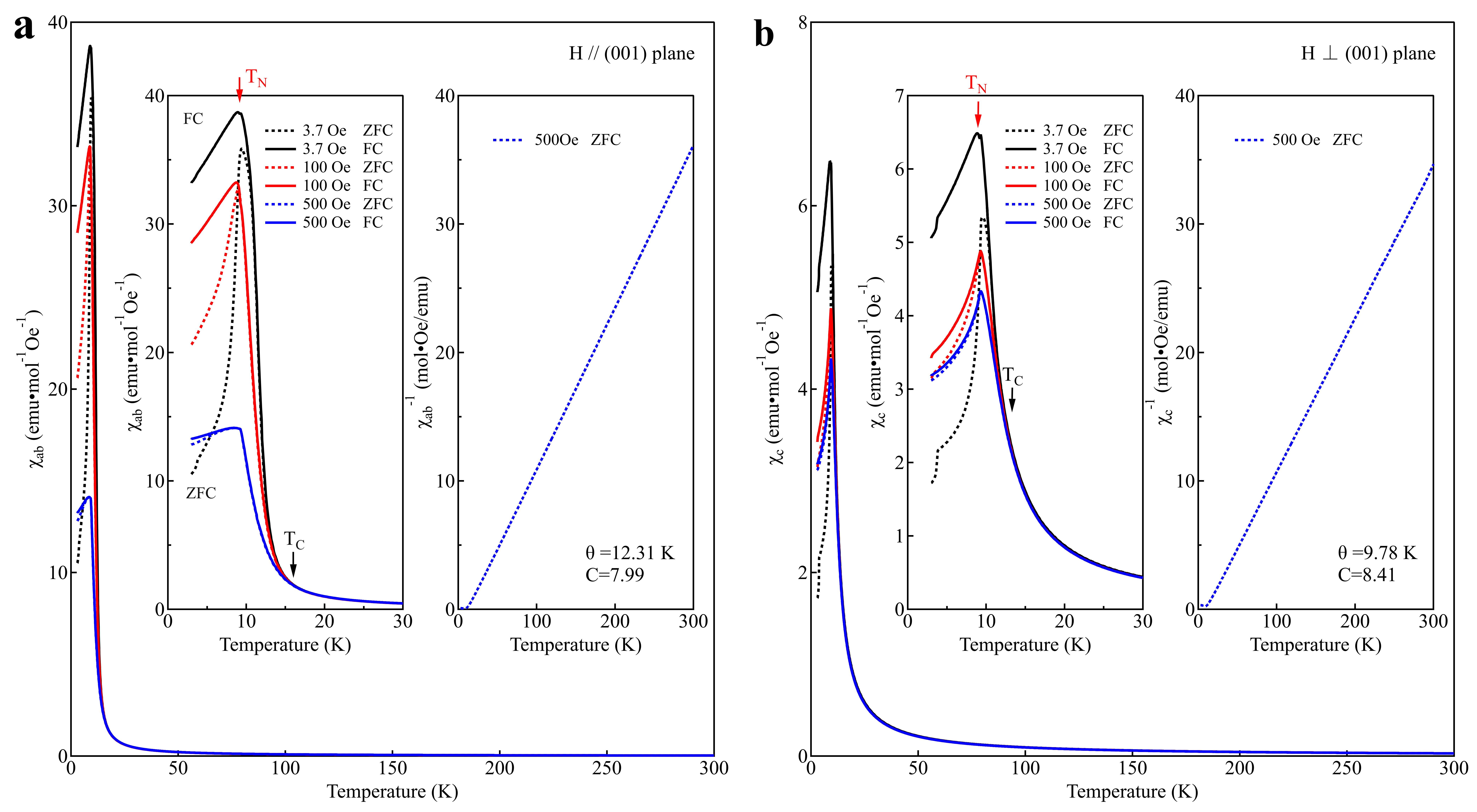}
\end{center}
\caption{\textbf{Magnetic susceptibility measurements of EuCd$_2$As$_2$.} (a-b) Temperature-dependent magnetic susceptibility measured with the applied magnetic field parallel (a) or perpendicular (b) to the (001) plane. They are measured at different external magnetic fields using field-cooled (FC) and zero-field-cooled (ZFC) modes. The upper-left insets show the zoom-in magnetic susceptibility measured at low temperature. The upper-right insets show the inverse magnetic susceptibility as a function of temperature. We note that there is a transition at $\sim$3.7\,K in some measurements which may be caused by the residual tin (superconducting with a T$_c$ at 3.7\,K) that was used during the growth of the EuCd$_2$As$_2$ single crystals by flux method.
}
\end{figure*}

\begin{figure*}[tbp]
\begin{center}
\includegraphics[width=1\columnwidth,angle=0]{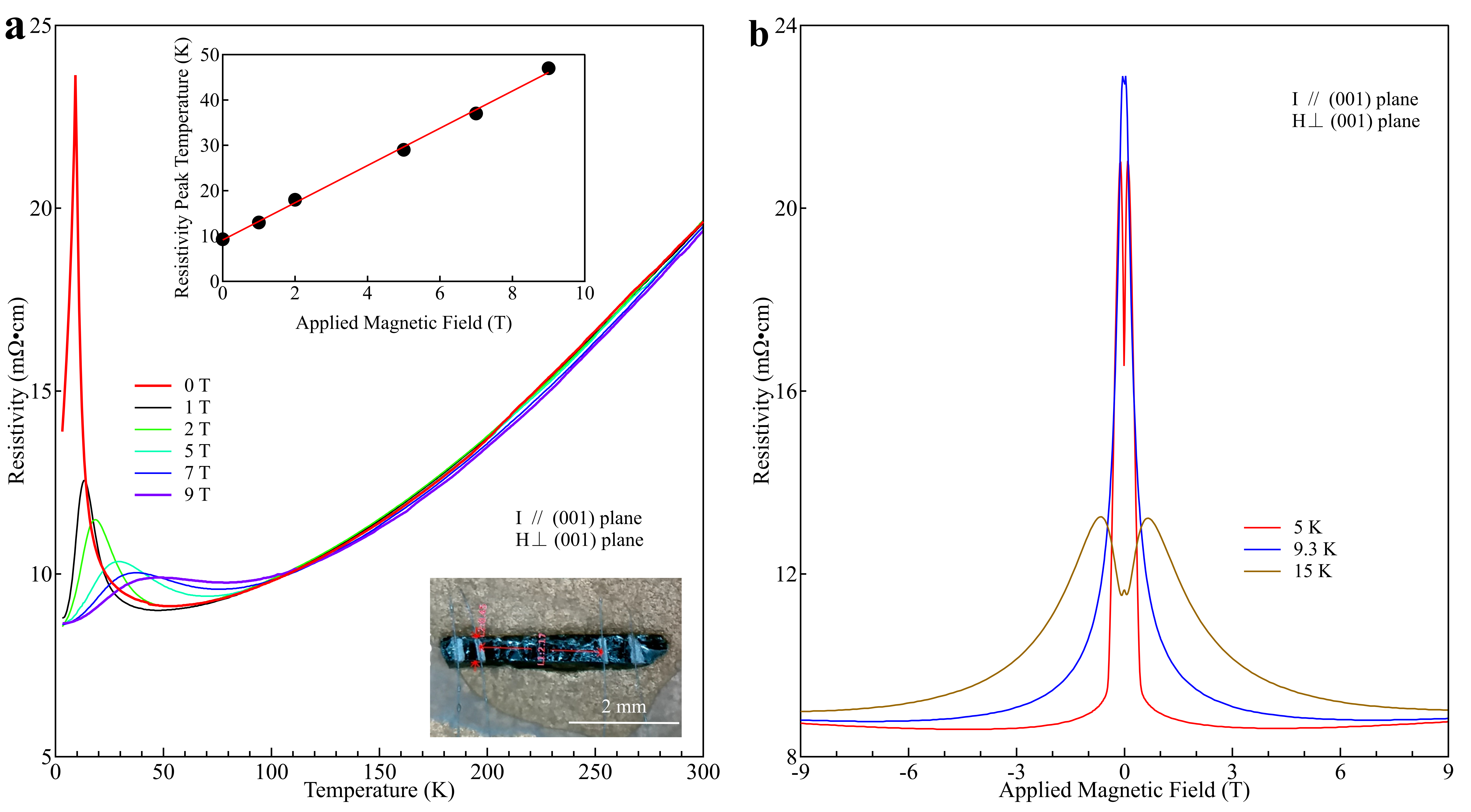}
\end{center}
\caption{\textbf{Resistivity measurements of EuCd$_2$As$_2$.} (a) Temperature-dependent resistivity measured on the (001) plane under different magnetic fields that are applied perpendicular to the (001) plane. The lower-right inset shows the measured sample with the four electrodes. The upper inset plots the resistivity peak temperature as a function of the applied magnetic field. (b) Magnetic field-dependent resistivity measured at, below and above the N\'eel temperature of 9.3\,K. The magnetic field is applied perpendicular to the (001) plane. 
}
\end{figure*}

\begin{figure*}[tbp]
\begin{center}
\includegraphics[width=1\columnwidth,angle=0]{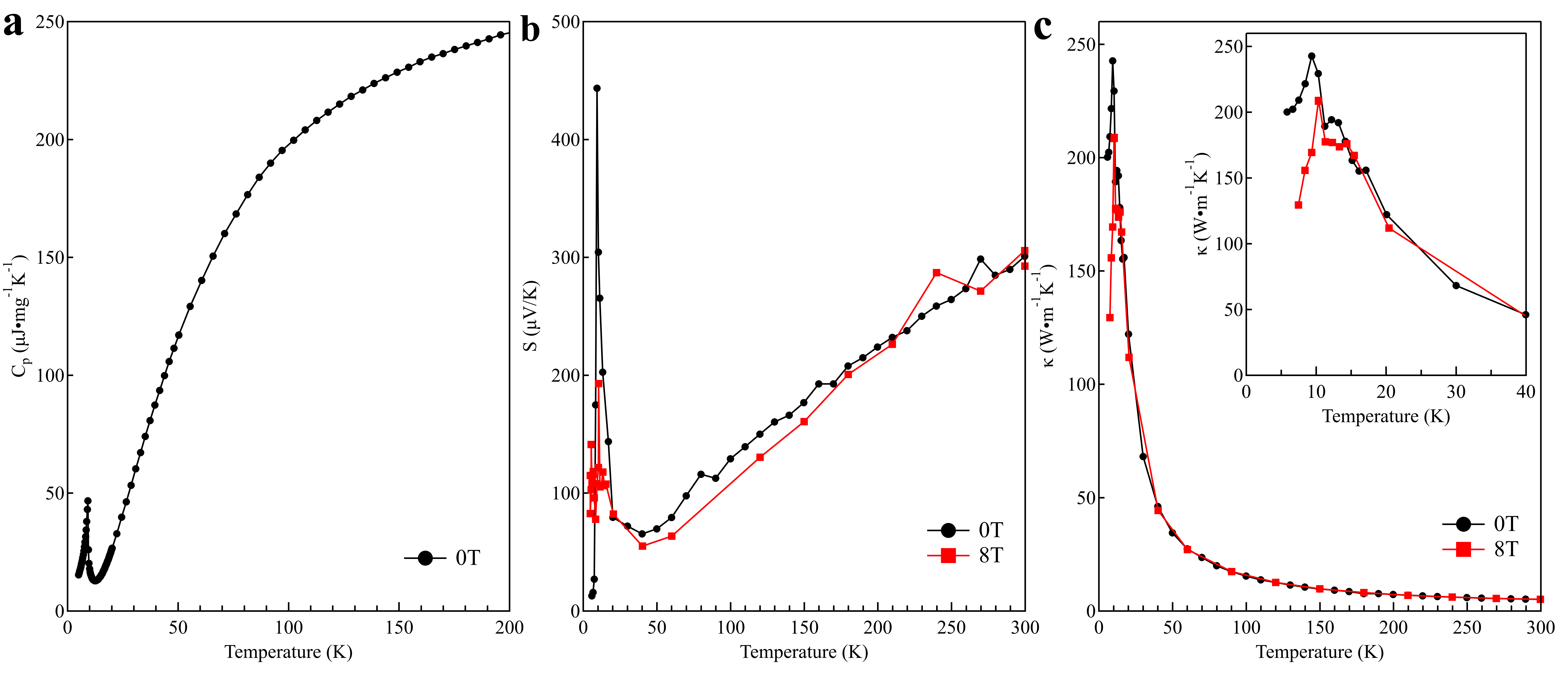}
\end{center}
\caption{\textbf{Thermodynamic measurements of EuCd$_2$As$_2$.} (a) Temperature dependence of the specific heat C$_p$ measured at zero magnetic field. (b) Temperature dependence of the Seebeck coefficient $S$ measured without (black curve) and with (red curve) magnetic field applied perpendicular to the (001) plane. (c) Temperature dependence of the thermal conductivity $\kappa$ measured without (black curve) and with (red curve) magnetic field applied perpendicular to the (001) plane.
}
\end{figure*}

\begin{figure*}[tbp]
\begin{center}
\includegraphics[width=1\columnwidth,angle=0]{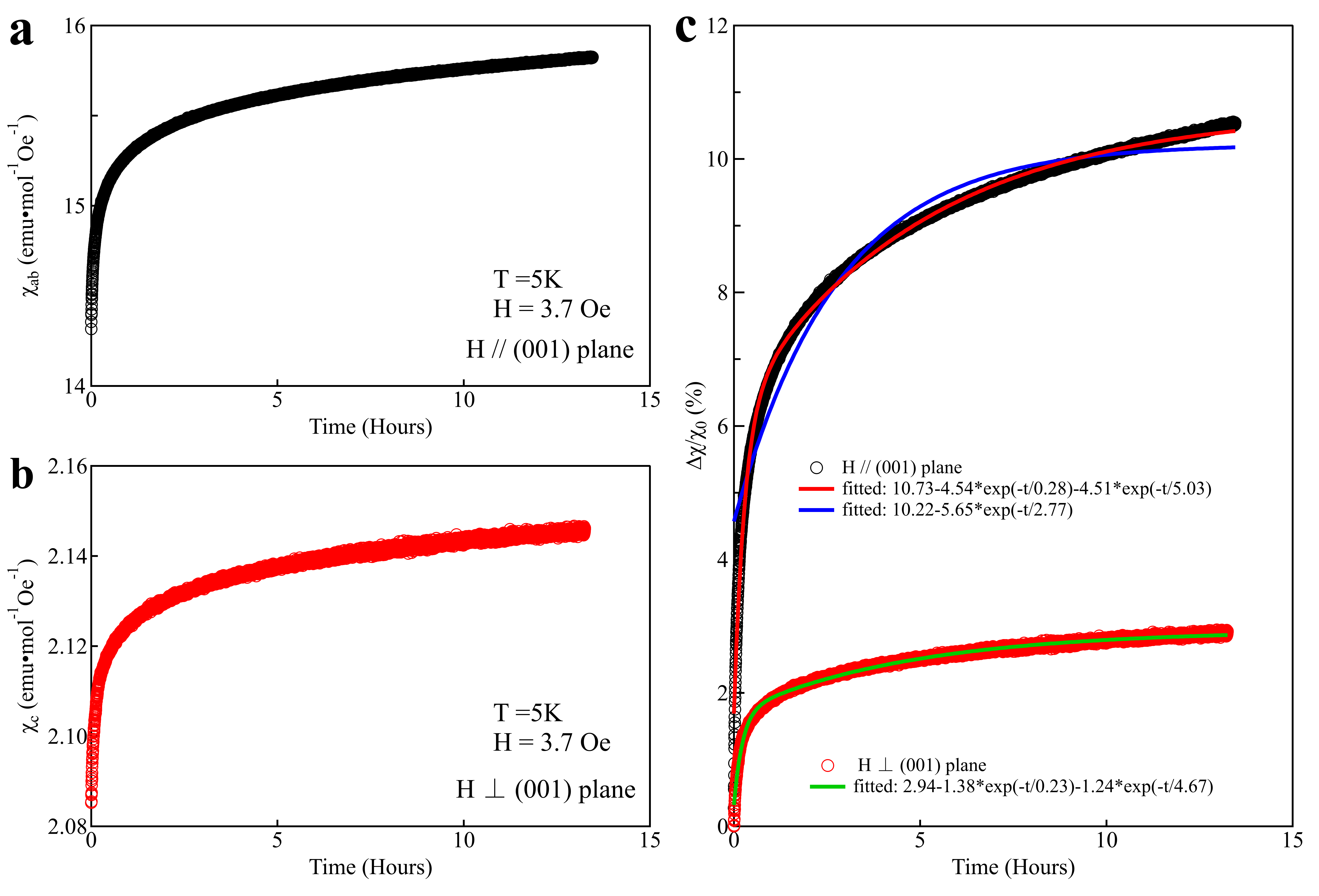}
\end{center}
\caption{\textbf{Slow magnetic relaxation in EuCd$_2$As$_2$.} (a) In-plane magnetic susceptibility as a function of the relaxation time measured after the sample is quickly cooled down from the room temperature to a low temperature of 5\,K. The applied magnetic field is 3.7\,Oe that is parallel to the (001) plane. (b) Out-of-plane magnetic susceptibility as a function of the relaxation time measured after the sample is quickly cooled down from the room temperature to a low temperature of 5\,K. The applied magnetic field is 3.7\,Oe and perpendicular to the (001) plane. (c) Comparison of the relative change between the in-plane and out-of-plane magnetic susceptibility as a function of the relaxation time. The measured data are fitted by two forms. The first one is $a+b*exp(-t/\tau)$ by considering only one relaxation time $\tau$. The second one is $a+b*exp(-t/\tau_1)+c*exp(-t/\tau_2)$ by considering two relaxation time $\tau_1$ and $\tau_2$.
}
\end{figure*}

\begin{figure*}[tbp]
\begin{center}
\includegraphics[width=1\columnwidth,angle=0]{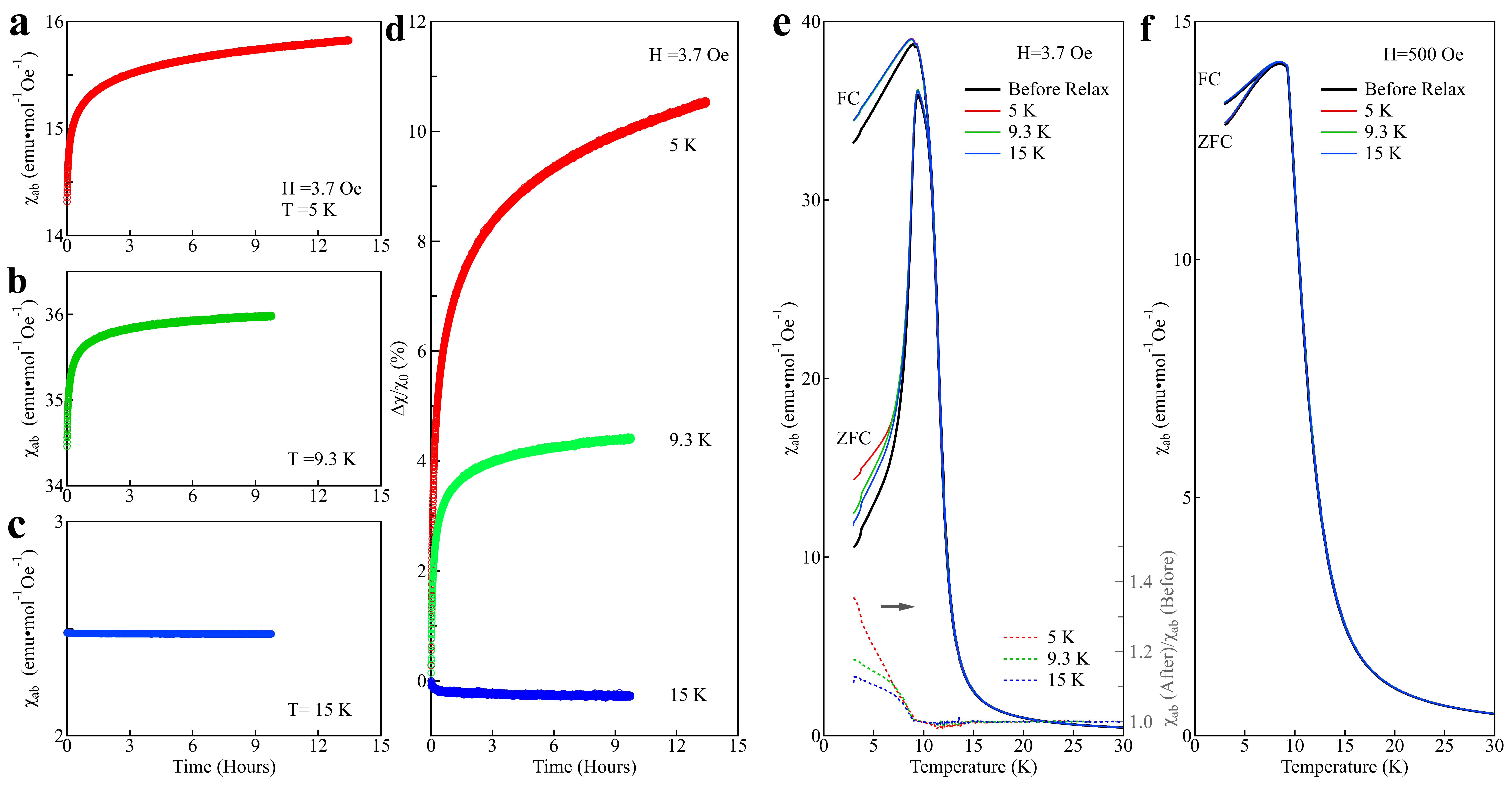}
\end{center}
\caption{\textbf{Magnetic relaxation at different temperatures in EuCd$_2$As$_2$.} (a-c) In-plane magnetic susceptibility as a function of the relaxation time measured after the sample is quickly cooled down from the room temperature to temperature of 5\,K (a), 9.3\,K (b) and 15\,K (c). The three measurements are independent and each measurement starts with the sample at the room temperature. The applied magnetic field is 3.7\,Oe and parallel to the (001) plane. (d) Comparison of the relative change of the in-plane magnetic susceptibility measured at different temperatures. The data are obtained from (a-c). (e-f) Temperature-dependent magnetic susceptibility of EuCd$_2$As$_2$ before and after the relaxation at different temperatures, measured with the applied magnetic field parallel to the (001) plane. They are measured at different external magnetic fields of 3.7\,Oe (e) and 500\,Oe (f) using both the field-cooled (FC, upper curves) and the zero-field-cooled (ZFC, lower curves) modes. The relative magnetic susceptibility change before and after the relaxation at different temperatures, $\chi_{ab}$(After)/$\chi_{ab}$(Before), is also plotted in (e) for the ZFC measurements.
}
\end{figure*}

\begin{figure*}[tbp]
\begin{center}
\includegraphics[width=1\columnwidth,angle=0]{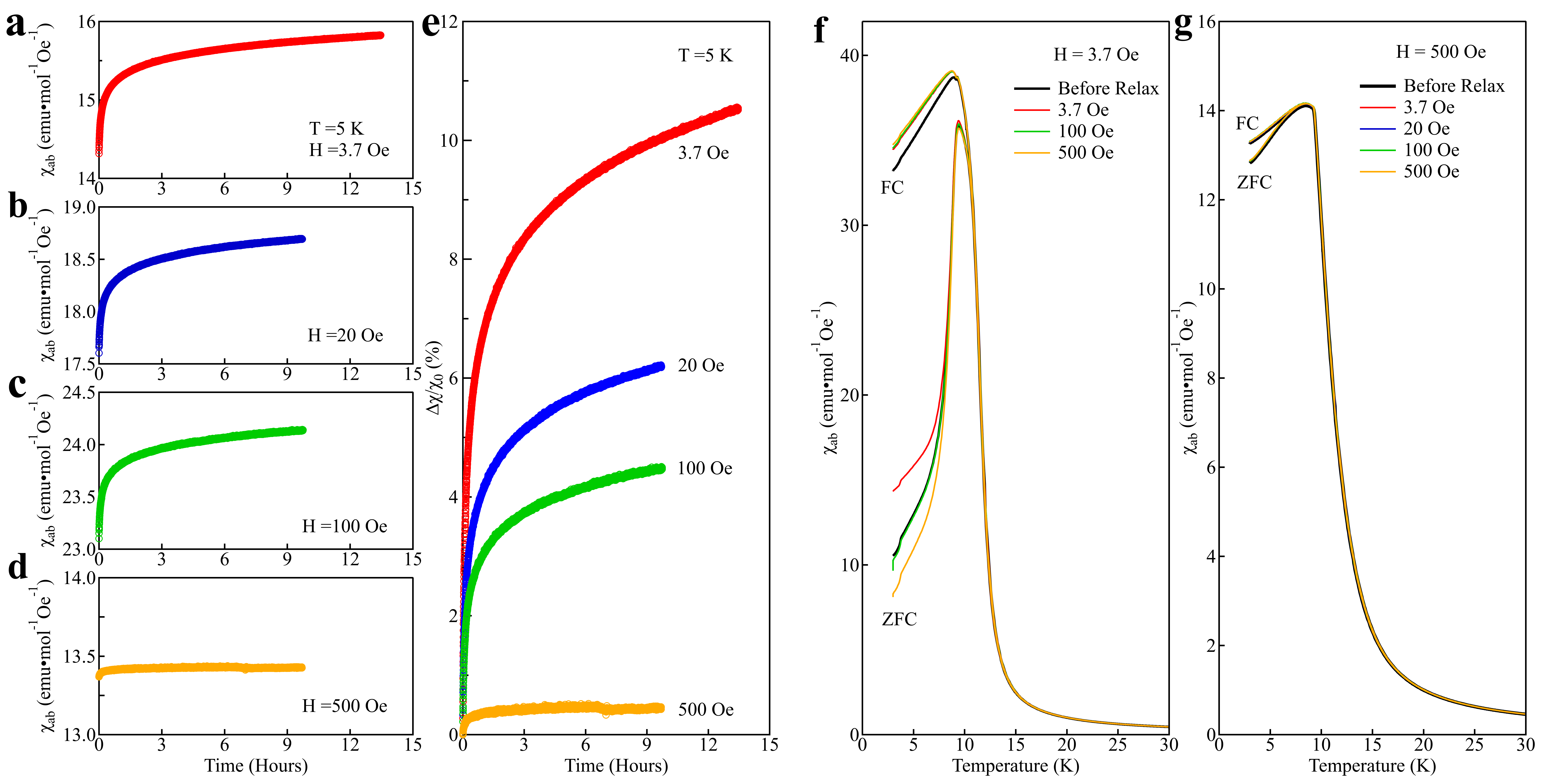}
\end{center}
\caption{\textbf{Magnetic relaxation under different magnetic fields in EuCd$_2$As$_2$.}(a) In-plane magnetic susceptibility as a function of the relaxation time. It is measured after the sample is quickly cooled down from the room temperature to a low temperature of 5\,K and then the magnetic field of 3.7\,Oe is applied parallel to the (001) plane and the magnetic susceptibility is measured at different relaxation time under the same magnetic field 3.7\,Oe. (b-d) Same as (a) but using different magnetic fields of 20\,Oe (b), 100\,Oe (c) and 500\,Oe (d), respectively. These measurements are independent and each measurement starts with the sample at the room temperature. (e) Comparison of the relative change of the in-plane magnetic susceptibility for the samples relaxed under different magnetic fields. The data are obtained from (a-d). (f-g) Temperature-dependent in-plane magnetic susceptibility of EuCd$_2$As$_2$ before and after the relaxation under different magnetic fields. They are measured at different external magnetic fields of 3.7\,Oe (f) and 500\,Oe (g) using both the field-cooled (FC, upper curves) and the zero-field-cooled (ZFC, lower curves) modes.
}
\end{figure*}

\begin{figure*}[tbp]
\begin{center}
\includegraphics[width=1\columnwidth,angle=0]{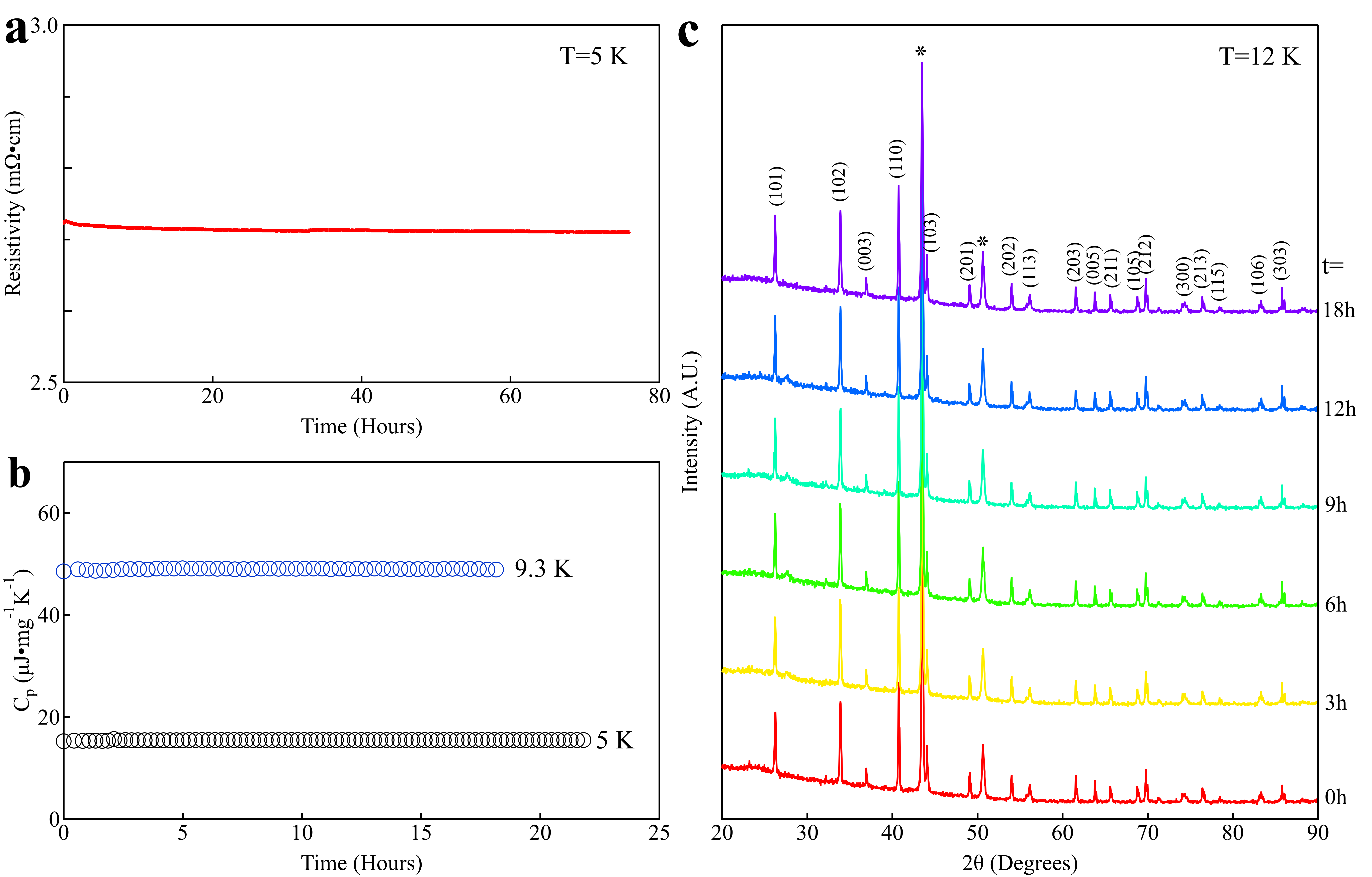}
\end{center}
\caption{\textbf{Time-dependent physical property and crystal structure measurements of EuCd$_2$As$_2$.} (a) Time-dependent measurement of the (001) in-plane resistivity measured at 5\,K. The sample was quickly cooled down from the room temperature to 5\,K before the resistivity was measured at different time. (b) Time-dependent measurement of the specific heat measured at 5\,K (black circles) and 9.3\,K (blue circles). The sample was quickly cooled down from the room temperature to 5\,K and then the specific heat was measured at different time. After this, the sample was warmed up to 200\,K and then cooled down quickly to 9.3\,K. The specific heat was measured at 9.3\,K at different time. (c) Time-dependent measurement of the powder XRD measured at 12\,K. The sample was quickly cooled down from the room temperature to 12\,K and the powder XRD pattern was measured at different time.
}
\end{figure*}

\end{document}